\documentclass[aps,prl,superscriptaddress,twocolumn]{revtex4}
\usepackage{amssymb}
\usepackage{graphicx}
\usepackage{amsmath}
\usepackage[colorlinks=true,linkcolor=blue,citecolor=blue,urlcolor=blue]{hyperref}

\begin{document}

\title{Universal Dynamics of a Degenerate Bose Gas Quenched to Unitarity}

\author{Chao Gao}
\email{gaochao@zjnu.edu.cn}
\affiliation{Department of Physics, Zhejiang Normal University, Jinhua, 321004, China}
\author{Mingyuan Sun}
\affiliation{Institute for Advanced Study, Tsinghua University, Beijing, 100084, China}
\author{Peng Zhang}
\affiliation{Department of Physics, Renmin University of China, Beijing, 100872,
China}
\affiliation{Beijing Key Laboratory of Opto-electronic Functional Materials \&
Micro-nano Devices, 100872 (Renmin Univeristy of China)}
\author{Hui Zhai}
\email{hzhai@tsinghua.edu.cn}
\affiliation{Institute for Advanced Study, Tsinghua University, Beijing, 100084, China}
\date{\today}

\begin{abstract}

Motivated by an unexpected experimental observation from the Cambridge group, [Eigen {\it et al.,} Nature {\bf 563}, 221 (2018)], we study the evolution of the momentum distribution of a degenerate Bose gas quenched from the weakly interacting regime to the unitary regime. 
For the two-body problem, we establish a relation that connects the momentum distribution at a long time to a sub-leading term in the initial wave function. For the many-body problem, we employ the time-dependent Bogoliubov variational wave function and find that, in certain momentum regimes, the momentum distribution at long times displays the same exponential behavior found by the experiment. 
Moreover, we find that this behavior is universal and is independent of the short-range details of the interaction potential. Consistent with the relation found in the two-body problem, we also numerically show that this exponential form is hidden in the same sub-leading term of the Bogoliubov wave function in the initial stages. Conceptually, our results show that, for quench to the universal regime and coherent quantum dynamics afterward, the universal long-time behavior is hidden in the initial state. 

\end{abstract}

\maketitle

Because the interaction in cold atomic systems can be controlled by optical and magnetic fields, it can be tuned in a time scale that is much shorter than the relaxation time. Cold atomic systems are also very clean and the microscopic interaction between atoms can be understood very well in terms of universal low-energy effective interactions. Because of these two reasons, cold atoms are ideal for studying far-from-equilibrium dynamics in the many-body sector from a microscopic point of view. In equilibrium, there exists a lot of phenomena that are universally applicable, independently of the details of interactions at the microscopic scale. A major question for the non-equilibrium physics is whether such universal phenomena can also be found in far-from-equilibrium situations.    

A recent experiment on strongly interacting Bose gases reveals a great surprise \cite{exp}. 
The system is initially prepared as a nearly pure Bose condensate at very low temperature and with weak interactions. 
Then the interaction is changed abruptly and the system is quenched to unitarity with the $s$-wave scattering length being infinite. The subsequent many-body dynamics was monitored by observing the evolution of the momentum distribution $n_{{\bf  k}}$. 
A prethermalization stage was found where $n_{{\bf  k}}$ remains a constant for a long time.
The most surprising finding in the experiment is that $n_{{\bf  k}}$ has a functional form
\begin{equation}
n_{{\bf  k}}\sim e^{-\Lambda k/k_n}, \label{nklong}
\end{equation}
where $k=|{\bf  k}|$, $k_n=(6\pi^2 n)^{1/3}$ is a momentum scale related to the total density $n$, and $\Lambda=3.62$ is obtained from fitting the experimental data \cite{exp}. This functional form is seen to be valid for $k$ ranging from $\sim k_n$ to a few times $k_n$. 
There are a number of previous theoretical works that have studied weakly interacting Bose gases quenched to the strongly interacting regime \cite{theory1,theory2,theory3,theory4,theory5,theory6,theory7,theory8,theory9,theory10,theory11}, with either finite or infinite scattering lengths. However, this phenomenon has not been predicted by any theory before. 

Here we should note that at unitarity, because there is no other length scale,
 both the two-body collisional rate and the three-body loss rate are proportional to $E_n$,
  where $E_n$ is given by $\hbar^2 k^2_n/(2m)$. 
However, it has been shown previously that the coefficient for the two-body collisional rate is usually larger than that for the three-body loss rate \cite{Rem13,Fletcher13,Jin,Eigen17}, such that the many-body dynamics is governed by two-body collisions for a reasonably long time before the three-body loss takes over and heats the system up. 
Therefore, it is very reasonable to view both the prethermalization and processes occurring before that as caused by the two-body collisions, while temperature increasing at later times is due to the three-body losses. 
This separation of time scales allows us to safely ignore the three-body loss and only focus on two-body collisional effects when analyzing the prethermal dynamics. In another words, we can view the prethermalization regime as the long time limit of the dynamics governed by the two-body collisions. Moreover, we have also ignored the coherent three-body effect like the Efimov effect, because usually when two-body interaction is dominated, the three-body effect only provides a small correction. 

In this letter we focus on understanding of the origin of the emergent exponential behavior of $n_{{\bf  k}}$ and answering whether this  functional form is universal or not, and we address this issue from both two-body and many-body perspectives. The main results can be summarized as follows:

{\bf  I.} For the two-body problem, we prove a relation between the long time behavior of the momentum distribution and the properties of the initial wave function. This relation works for arbitrary short-range potentials. With this theorem, we can determine which property of the initial wave function is responsible for the exponential form of Eq.~(\ref{nklong}) in the momentum distribution at a long time.   

{\bf  II.} For the many-body problem, we employ a variational time-dependent Bogoliubov wave function and by solving the time-dependent equation, we find that for a certain range of momentum, the averaged $n_{{\bf  k}}$ for long time evolution indeed obeys the form of Eq.~(\ref{nklong}), although the coefficient $\Lambda$ is quantitatively different from the experimental value due to the mean-field nature of our ansatz. We use three different potentials tuned near the vicinity of a scattering resonance, that are the square well, the Gaussian potential and the Yukawa potential, and find that this behavior is independent of the short-range details. 

At the end, we also discuss the connection between the two-body and the many-body results. 

\textit{Two-Body Problem.} Let us first start with the two-body problem whose Hamiltonian can be written in terms of the relative coordinate ${\bf  r}$ as ${\hat H}={\hat H}_0+{\hat V}({\bf  r})$, where $\hat{H}_0=-\hbar^2\nabla^2/m$ is the kinetic energy with $m$ being the mass, and $\hat{V}({\bf  r})$ is a short-range potential and we only consider the $s$-wave interaction. Here we choose $t=0$ as the time right after the quench of interactions and we denote the initial wave function as $|\phi^{i}\rangle$. The momentum distribution $n_{{\bf  k}}(t)$ at momentum ${\bf  k}$ and time $t$ is given by 
\begin{equation}
n_{{\bf  k}}(t)=|\langle {\bf  k}|e^{-\frac{i}{\hbar}\hat{H}t}|\phi^{i}\rangle|^2=|\langle {\bf  k}|e^{\frac{i}{\hbar}\hat{H}_0 t}e^{-\frac{i}{\hbar}\hat{H}t}|\phi^{i}\rangle|^2,
\end{equation}
where $|{\bf  k}\rangle$ is a plane wave state. The second equality follows from the fact that $|{\bf  k}\rangle$ is an eigenstate of $\hat{H}_0$ and $e^{\frac{i}{\hbar}\hat{H}_0t}$ only gives rise to a phase factor that does not change $n_{{\bf  k}}$. Furthermore, making use of the properties of the M{\o}ller operator \cite{Taylor}
\begin{equation}
\hat{\Omega}^{(-)}=\lim\limits_{t\rightarrow +\infty}e^{\frac{i}{\hbar}\hat{H} t}e^{-\frac{i}{\hbar}\hat{H}_0t}
\end{equation}
in scattering theory, we can derive the following relation \cite{supple}
\begin{equation}
n_{{\bf  k}}(t\rightarrow +\infty)=|\langle {\bf  k}|\hat{\Omega}^{(-)\dagger}|\phi^{i}\rangle|^2=|\langle {\bf  k}^{(-)}|\phi^{i}\rangle|^2. \label{theorem}
\end{equation}
Here $|{\bf  k}^{(-)}\rangle$ is the inward scattering wave function defined as \cite{Taylor}
\begin{equation}
|{\bf  k}^{(-)}\rangle=|{\bf  k}\rangle+\frac{1}{\epsilon_{{\bf  k}}+i0^--{\hat{H}}}\hat{V}|{\bf  k}\rangle,
\end{equation}
where $\epsilon_{{\bf k}}=\hbar^2 {\bf  k}^2/m$. For a short-range potential, it is straightforward to show that outside the range of interaction, $\langle {\bf  r}|{\bf  k}^{(-)}\rangle$ behaves as \cite{Taylor}
\begin{equation}
\langle {\bf  r}|{\bf  k}^{(-)}\rangle=\frac{1}{(2\pi)^{3/2}}\left(e^{i{\bf  k}\cdot{\bf  r}}+\frac{1}{ik}\frac{e^{-ikr}}{r}\right). \label{k-r}
\end{equation}
where $r=|{\bf  r}|$ and $k=|{\bf k}|$. 
Here we have explicitly used the fact that the system is quenched to unitarity with $a_\text{s}=\infty$. 

With the relation Eq.~(\ref{theorem}), we can determine the requirement for the initial wave function $|\phi^{i}\rangle$ that can lead to the long-time behavior of Eq.~(\ref{nklong}) in $n_{{\bf  k}}$. It is important to know that \{$|{\bf  k}^{(-)}\rangle$\} also form a complete and orthogonal basis, and we can expand the wave function in terms of this basis. Let us introduce 
\begin{equation}
\psi({\bf  k})=e^{i\theta({\bf  k})}\sqrt{n_{{\bf  k}}(t\rightarrow+\infty)}, \label{psik1}
\end{equation}
then the exponential form of $n_{{\bf  k}}$ will translate to the same kind of exponential dependence for $\psi({\bf  k})$ up to a phase factor. 
We can then write the initial wave function $|\phi^{i}\rangle$ as 
\begin{equation}
|\phi^{i}\rangle=\int d^3{\bf  p}\psi({\bf  p})|{\bf  p}^{(-)}\rangle,
\end{equation}
and in the momentum space
\begin{equation}
\phi^i({\bf  k})=\langle {\bf k}|\phi^{i}\rangle=\frac{1}{(2\pi)^{\frac32}}\int d^3{\bf  r}d^3{\bf  p} e^{-i{\bf  k}\cdot{\bf  r}}\psi({\bf  p})\langle {\bf  r}|{\bf  p}^{(-)}\rangle.  \label{psik}
\end{equation}

Considering the situation where $\psi({\bf  p})$ is isotropic, i.e. it can be written as $\psi(p)$ with $p=|{\bf  p}|$, we can substitute Eq.~(\ref{k-r}) into Eq.~(\ref{psik}) and integrate out the azimuthal degrees of freedom, we find \cite{supple}
\begin{align}
&\phi^i(k)
=\frac{1}{2\pi }\left(-\frac{i}{k}\right)\lim_{\epsilon \to 0^+} \int _0^{+\infty }dp p \psi(p) \nonumber
\\
&\times\left[\sum\limits_{\sigma=\pm}\frac{\sigma}{p+\sigma(k+i\epsilon )}  
+\sum\limits_{\sigma^\prime=\pm}\frac{\sigma'}{p+\sigma^\prime(k-i \epsilon )}\right]. \label{phik}
\end{align}
Let us introduce an auxiliary function $\Psi(z)$ in the complex plane, such that it satisfies the requirement at the positive side of the real axis $\Psi(z=p>0)=\psi(p)$ and at its negative side $\Psi(z=p<0)=-\psi(-p)$ \cite{note}. With the help of this auxiliary function, it can be shown that 
\begin{equation}
\phi^{i}(k)=-\psi(k)-\frac{1}{k}\sum\limits_{j}\frac{\text{Res}\left[2z\Psi(z)\right]_{z=z_j}}{z_j-k}, \label{psik2}
\end{equation}
where $\text{Res}[f(z)]_{z=z_j}$ denotes the residue of the function $f(z)$ at its pole $z_j$.

Eq.~(\ref{psik2}) is a very interesting result. Here we should note that the amplitude of $\psi(k)$ obeys this universal exponential form only at the momentum $\gtrsim k_n$, however, the auxiliary function $\Psi(z)$ will certainly depend on the small momentum behavior of $\psi(k)$. Therefore, the residues of $\Psi(z)$, as well as the coefficient for this second term in the r.h.s. of Eq.~(\ref{psik2}), are non-universal. When $\psi(k)$ is a regular function as in Eq.~(\ref{nklong}) and Eq.~(\ref{psik1}), the second term recovers the well-known $1/k^4$ behavior of the momentum distribution at large $k$. 
Hence, Eq.~(\ref{psik2}) tells us that, one can subtract the leading order $1/k^2$ term from fitting the large momentum, and the remaining regular sub-leading term reveals the momentum distribution at a long time. 
That is to say that, in order for the long time behavior of $n_{{\bf  k}}$ to obey Eq.~(\ref{nklong}),
 the sub-leading term of the initial wave function has to obey the form given by Eq.~(\ref{nklong}) and Eq.~(\ref{psik1}).

\begin{figure}[t]
\includegraphics[width=0.9\linewidth]{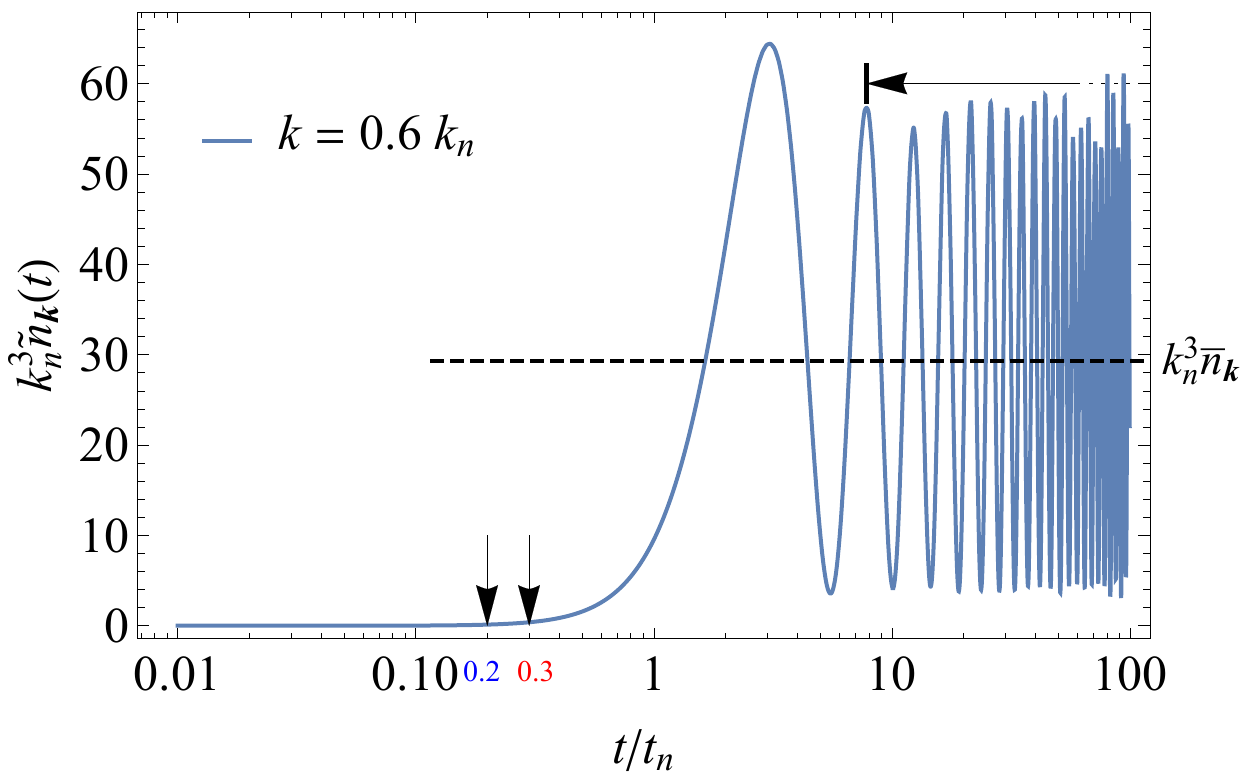} 
\caption{
 A typical value of the normalized $\tilde{n}_{{\bf  k}}(t)$ (in unit of $1/k^3_n$) is plotted as a function of $t$. Here we take $k=0.6k_n$ and the interaction potential is the Yukawa potential. The horizontal line with arrow indicates the time domain in which we take average of $\tilde{n}_{{\bf  k}}(t)$ to obtain $\bar{n}_{{\bf  k}}$. The two vertical arrows indicates two time slots where the wave function is plotted in Fig. \ref{gksub}. \label{nkt} }
\end{figure}

\textit{Many-Body Problem.} Now we turn into the many-body problem whose Hamiltonian can be written in second quantized form as
\begin{equation}
\hat{\mathcal{H}}=\sum _{{\bf k}}\epsilon _{{\bf k}}\hat{a}_{{\bf k}}^{\dagger }\hat{a}_{{\bf k}}+\frac{1}{2L^3}\sum _{{\bf k},{\bf k}',{\bf q}} \hat{a}_{{\bf k}+{\bf q}}^{\dagger}\hat{a}_{{\bf k}'-{\bf q}}^{\dagger }V({\bf q}) \hat{a}_{{\bf k}'} \hat{a}_{{\bf k}}.
\end{equation}
Here $\hat{a}_{{\bf k}}^{\dagger }$($\hat{a}_{{\bf k}}$) is the creation (annihilation) operator for bosons with momentum ${\bf  k}$. $L^3$ is the system's volume. $V({\bf q})=\int d^3{\bf r}e^{i{\bf q} {\bf r}}V({\bf r})$ is the Fourier transform of the interaction potential $V({\bf  r})$. We implement the Bogoliubov-type variational ansatz
\begin{equation}
|\Phi (t)\rangle =\mathcal{A}(t)\exp \left[g_0(t)\hat{a}_0^{\dagger }+\sum _{{\bf k}\cdot \hat{z}>0} g_{{\bf k}}(t)\hat{a}_{{\bf k}}^{\dagger
}\hat{a}_{-{\bf k}}^{\dagger }\right]|0\rangle . \label{Eq:ansatz}
\end{equation}
Here $\mathcal{A}(t)$ is a normalization factor, $|0\rangle$ is the particle vacuum; $g_0$ and $g_{{\bf k}}$ are variational parameters, which can determine $N_0(t)=|g_0|^2$ and $N_{{\bf  k}}(t)=|g_{{\bf k}}|^2/(1-|g_{{\bf k}}|^2)$ as the particle number at zero-momentum and finite momentum ${\bf  k}$ modes. The Bogoliubov ansatz assumes that the system remains as a Bose condensate during the entire dynamics, which is indeed the case for this experiment. 

The dynamical equations for $g_0(t)$ and $g_{{\bf  k}}(t)$ can be obtained from the Euler-Lagrange equation for the Lagrangian  $\mathcal{L}=\frac{i\hbar}{2}[\langle \Phi|\dot{\Phi }\rangle -\langle \dot{\Phi } |\Phi \rangle ]-\langle \Phi| \hat{H}|\Phi\rangle$. 
It results in coupled equations for $g_0$ and $g_{{\bf  k}}$ as \cite{theory2}
\begin{align}
i\hbar\dot{g}_0&=n V(0)g_0+\frac{1}{L^3}\sum _{{\bf k}\neq {\bf 0}} V({\bf k})\frac{g_0^*g_{{\bf k}}+g_0| g_{{\bf k}}|^2}{1-|
g_{{\bf k}}|^2},\nonumber
\\
i\hbar\dot{g}_{{\bf p}}&=2[\epsilon _{{\bf p}}+n V(0)]g_{{\bf p}}+\frac{V({\bf p})}{L^3}[g_0^2+g_0^{*2}g_{{\bf p}}^2+2|
g_0| {}^2g_{{\bf p}}]\nonumber
\\
&+\frac{1}{L^3}\sum _{{\bf k}\neq {\bf 0}}V({\bf p}-{\bf k})\frac{2|g_{{\bf k}}|
{}^2g_{{\bf p}}+g_{{\bf k}}+g_{{\bf k}}^*g_{{\bf p}}^2}{1-|g_{{\bf k}}|^2}.\label{eq:EoM}
\end{align}
The total number $N=N_0+\sum_{{\bf  k}\neq 0}N_{{\bf  k}}(t)$ is a conserved quantity, and $n=N/L^3$ is the total density. Making use of the spherical symmetry of this system, we consider that $g_{{\bf  k}}$ only depends on $k$ and can be simplified as $g_k$, and we can further simplify this equation by performing the azimuthal integration first \cite{supple}. Without loss of generality, we take the initial state to be a pure Bose-Einstein Condensate (BEC), i.e. $g_0(0)=\sqrt{N}$ and all $g_{{\bf k}}(0)=0$. As in the two-body case, we start the time evolution right after the interaction quench, and therefore we set the interaction at scattering resonance.

\begin{figure}[t]
\includegraphics[width=0.9\linewidth]{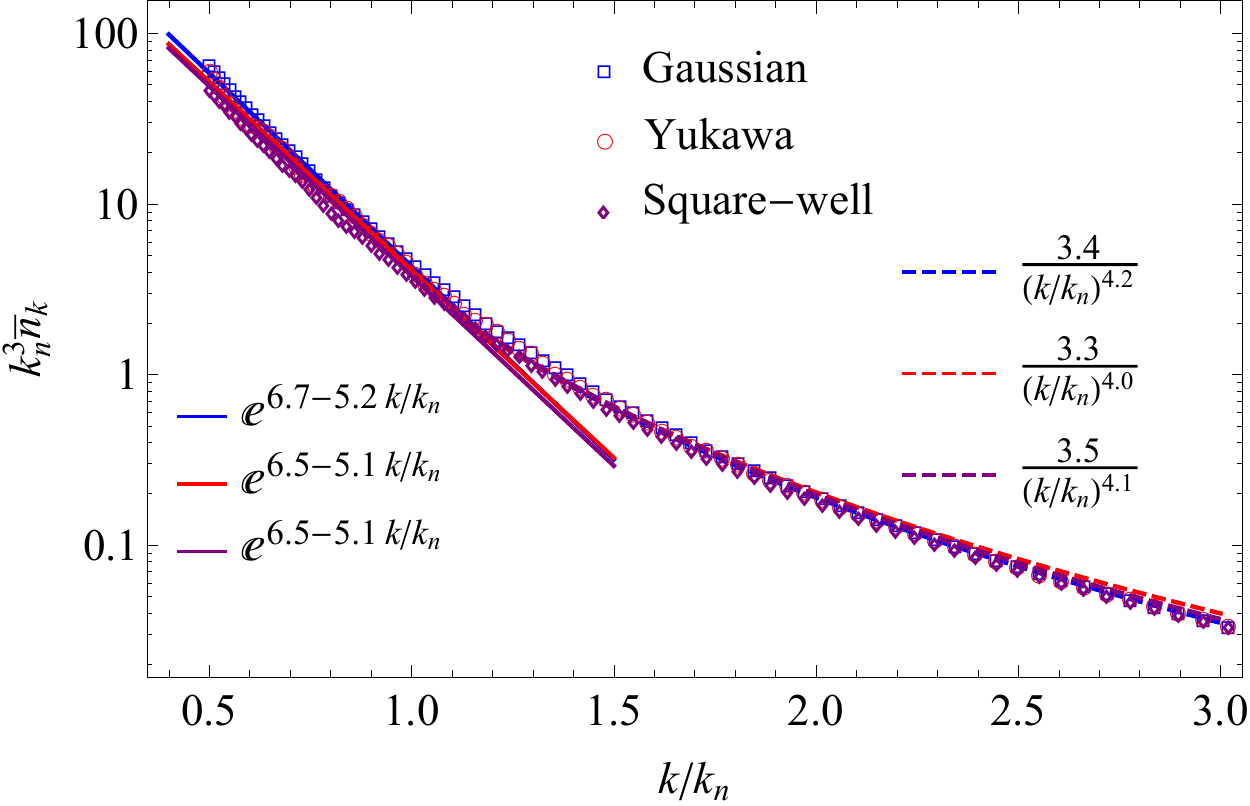} 
\caption{(color online)
$\bar{n}_{{\bf  k}}$ (in unit of $1/k^3_n$) is plotted as a function of $k/k_n$. Three different microscopic potentials are used in the calculation. \label{fitting} }
\end{figure}

Here, to verify whether the dynamics is universal, that is to say, whether it depends on the short-range details, we consider three different short-range potentials:

(i) The square well potential: \\$V_{\text{SW}}({\bf r})=-\frac{\hbar^2\gamma_\text{s}}{mr_0^2}\Theta \left(r_0-r\right)$, where $\Theta$ is the Heaviside step function. The $s$-wave resonance occurs at $\gamma_\text{s}=(\pi/2)^2$.

(ii) The Gaussian potential: \\
$V_{\text{GW}}({\bf r})=-\frac{\hbar^2\gamma _\text{g}}{mr^2_0}e^{-r^2/r_0^2}$, and the $s$-wave resonance occurs at $\gamma_\text{g}\approx2.68$.

(iii) The Yukawa potential:\\
 $V_{\text{YW}}({\bf r})=-\frac{\hbar^2\gamma_\text{y}}{mr_0} \frac{e^{-r\left/r_0\right.}}{4\pi r}$, and the $s$-wave resonance occurs at $\gamma_\text{y}\approx21.1$.

We numerically solve the coupled equations Eq.~(\ref{eq:EoM}) with these three potentials by discretizing both the radial momentum and the time, from which we can obtain $g_{{\bf  k}}(t)$ and $N_{{\bf  k}}(t)$. Following Ref. \cite{exp}, we introduce a normalized momentum distribution  
\begin{equation}
\tilde{n}_{{\bf  k}}(t)=\frac{N_{\bf  k}(t)}{n},
\end{equation}
such that $\frac{1}{L^3}\sum _{{\bf  k}} \tilde{n}_{{\bf  k}}(t)=1$.
In Fig. \ref{nkt} we plot $\tilde{n}_{{\bf  k}}(t)$ as a function of $t$. One can see that following a growth at the initial stage, $\tilde{n}_{{\bf  k}}(t)$ exhibits an oscillatory behavior for $t\gg t_n$, where $t_n$ is a typical time scale defined as $t_n=\hbar/E_n$. We also find that this oscillatory solution is stable against noises. Though it looks surprising that the highly non-linear equations can display stable oscillatory solution, it can be understood analytically in term of a simplified version of these coupled equations \cite{supple}. In reality, the interactions between quasi-particles cause Baliaev-Landau damping, which will eventually smear out the oscillation and lead to a saturation result. Here, we take a long time average of $\tilde{n}_{{\bf  k}}(t)$ starting from the second peak in the oscillation, as indicated in Fig. \ref{nkt}. The average is denoted by $\bar{n}_{{\bf  k}}$, which is taken as the long time saturation value of the momentum distribution.

In Fig. \ref{fitting}, we plot the dimensionless quantity $k_n^3\bar{n}_{{\bf  k}}$ as a function of $k/k_n$. The fit shows a regime around $k\sim k_n$, where $\bar{n}_{{\bf  k}}$ behaves as Eq.~(\ref{nklong}), consistent with the experimental observation in Ref. \cite{exp}. This fitting yields a coefficient $\Lambda=5.1-5.2$. For large $k$, the fitting yields a $1/k^4$ behavior. 
Most importantly, we note that 
the curves obtained using the three different potentials defined above collapse onto one another,
 which shows that this emergent exponential behavior of
 the momentum distribution is independent of the short-range details of the interaction potentials.

\begin{figure}[t]
\includegraphics[width=0.9\linewidth]{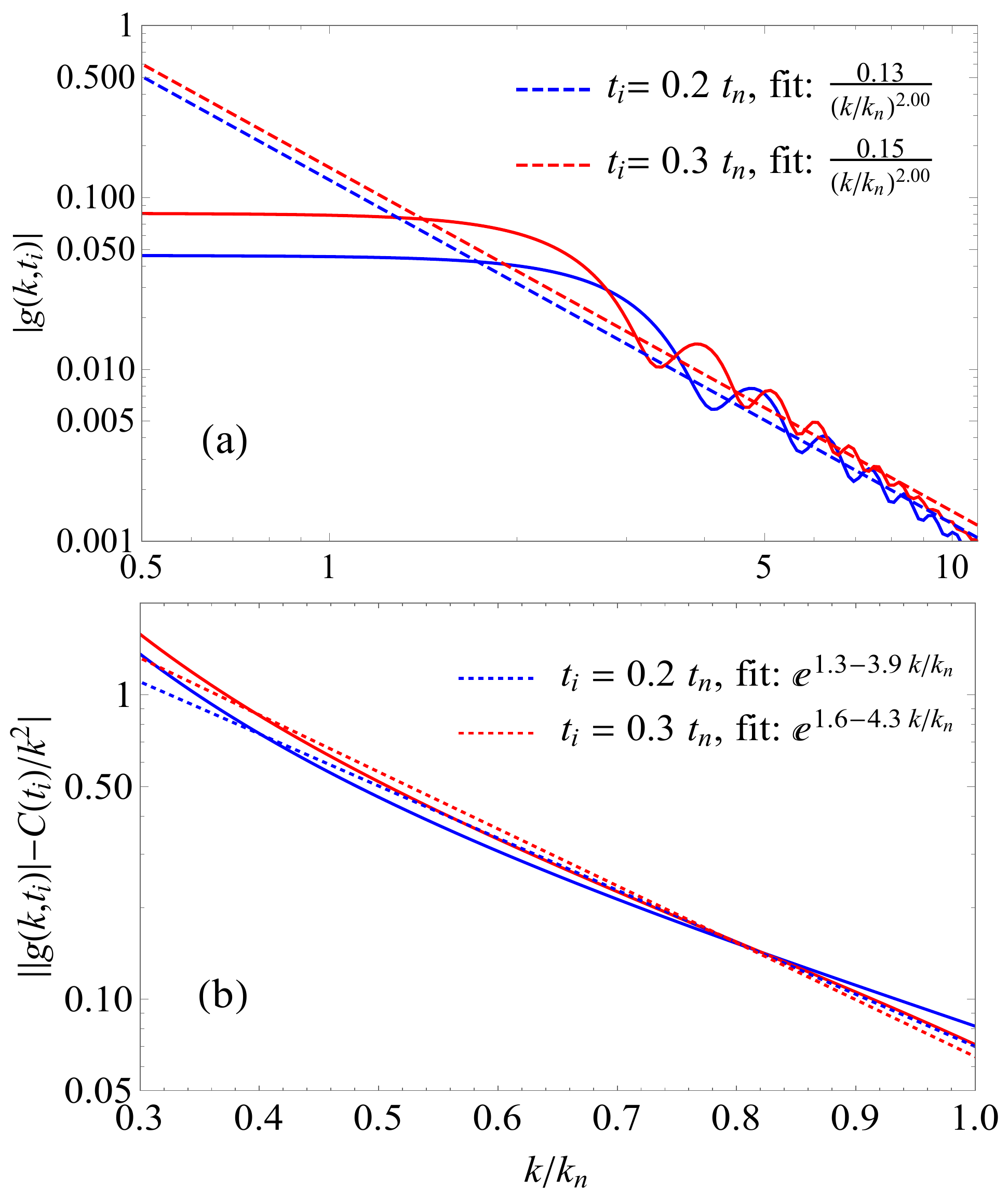} 
\caption{(color online) The momentum space Bogoliubov wave function $g_{{\bf  k}}$ for two time slots at the early time $t=0.2 t_n$ (blue lines) and $t=0.3 t_n$ (red lines) as marked in Fig. \ref{nkt}. In (a), the log-log plot shows that the large $k$ part of $|g_{k}|$ can be well fitted by $\sim 1/k^2$. In (b), the log plot shows that for the intermediate $k\sim k_n$, after subtracting $1/k^2$ part, the sub-leading term can be well fitted by $\sim e^{-\Lambda k/k_n}$.
\label{gksub} }
\end{figure}

\textit{Connection between the Two- and Many-Body Problems.} 
To summarize the results above, on one hand, our discussion on the two-body problem has established a relation between $n_{{\bf  k}}$ at the long time and the wave function behavior at the initial time; and on the other hand, our Bogoliubov calculation for the many-body problem has discovered the exponential form for $n_{{\bf  k}}$ at the long time as observed in the experiment reported in Ref.~\cite{exp}. Now a natural question is whether the same relation also holds in the Bogoliubov wave function, namely, whether the exponential form is also hidden in the sub-leading term of the Bogoliubov wave function at the early time. 
To check this conjecture, we look into the wave function $g_{k}$ at times $t<t_n$, far before the saturation of the momentum distribution, as indicated by arrows in Fig. \ref{nkt}. This early stage wave function is reminiscent of the initial wave function in the two-body case. In Fig. \ref{gksub}(a), we plot $|g_{k}|$ as a function of $k/k_n$, which does not show any exponential behavior,  and the large $k$ part can be well fitted by a $1/k^2$ tail. In Fig. \ref{gksub}(b), following the same spirt of Eq.~(\ref{psik2}) discovered in the two-body problem, we subtract the $1/k^2$ part in $|g_{k}|$, and plot the sub-leading term as a function of $k/k_n$. Interestingly, in the same momentum range where the long time $\bar{n}_{{\bf  k}}$ plotted in Fig. \ref{fitting} shows an exponential behavior, this sub-leading term in the early time wave function plotted in Fig. \ref{gksub}(b) also shows an exponential behavior. Notice that we start from an initial state with all atoms in the zero-momentum state, the exponential behavior at the early stage wave function may originate from the pair production process, as discussed in Ref. \cite{Cheng}.

\textit{Comments on Comparison with Experiment.} Aside from momentum distribution, we find that the growth time also follows the scaling law as discovered by the experiment, and we find that the condensed fraction at long time is not vanishing and is about $10\%$. However, we should emphasize that the agreement between our Bogoliubov theory and the experiment is only qualitative. Experimentally, this exponential behavior of $n_{{\bf  k}}$ is valid up to $\sim 3k_n$ and they do not find $1/k^4$ behavior, but in our case it is only up to $\sim k_n$ and is followed by a $1/k^4$ tail at higher momenta. The value of $\Lambda$ is also somewhat different between our calculation and the experimental result. 
However, since the system is strongly interacting, we do not expect the mean-field type Bogoliubov theory to be quantitatively accurate anyway. Moreover, our calculation leads to a very fast oscillation of $n_{{\bf  k}}$ at long times and its mean value saturates. In experiment, the momentum distribution eventually takes off again after a prethermalization plateau, and this is caused by the heating due to the three-body loss which we do not include in our theory. 
Our results offer valuable insight for understanding this observation but more involved theories are required for a more quantitative comparison with experiment.  

\textit{Note Added.} When finishing this paper, we became aware of another preprint, Ref. \cite{Meera}, which also did the many-body calculation with the Bogoliubov wave function. 

\textit{Acknowledgment.} We thank Wei Zheng, Manuel Valiente, Xin Chen, Zeng-Qiang Yu, Shizhong Zhang, Ran Qi for inspiring discussion.
We are particularly grateful to Manuel Valiente for carefully reading our manuscript and helpful suggestions. 
This work is supported by NSFC Grant No. 11734010, 11604300, 11835011, 11774315, Beijing Outstanding Young Scholar Program and MOST under Grant No. 2016YFA0301600.

\bibliographystyle{unsrt}

\begin{thebibliography}{99}

\bibitem{exp}
C. Eigen, J. A. Glidden, R. Lopes, E. A. Cornell, R. P. Smith, and Z. Hadzibabic, Nature {\bf 563}, 221 (2018).

\bibitem{theory1}
X. Yin and L. Radzihovsky, Phys. Rev. A {\bf 88}, 063611 (2013).
\bibitem{theory2}
A. G. Sykes, J. P. Corson, J. P. D'Incao, A. P. Koller, C. H. Greene, A. M. Rey, K. R. Hazzard, and J. L. Bohn, Phys. Rev. A {\bf 89}, 021601 (2014).
\bibitem{theory3}
A. Ran{\c{c}}on and K. Levin, Phys. Rev. A {\bf 90}, 021602 (2014).
\bibitem{theory4}
B. Kain and H. Y. Ling, Phys. Rev. A {\bf 90}, 063626 (2014).
\bibitem{theory5}
J. P. Corson and J. L. Bohn, Phys. Rev. A {\bf 91}, 013616 (2015).
\bibitem{theory6}
F. Ancilotto, M. Rossi, L. Salasnich, and F. Toigo, Few-Body Syst. {\bf 56}, 801 (2015).
\bibitem{theory7}
X. Yin and L. Radzihovsky, Phys. Rev. A {\bf 93}, 033653 (2016).
\bibitem{theory8}
V. E. Colussi, J. P. Corson, and J. P. D'Incao, Phys. Rev. Lett. {\bf 120}, 100401 (2018).
\bibitem{theory9}
V. E. Colussi, S. Musolino, and S. J. J. M. F. Kokkelmans, Phys. Rev. A {\bf 98}, 051601 (2018).
\bibitem{theory10}
M. Van Regemortel, H. Kurkjian, M. Wouters, and I. Carusotto, Phys. Rev. A {\bf 98}, 053612 (2018).
\bibitem{theory11}
J. P. D'Incao, J. Wang, and V. E. Colussi, Phys. Rev. Lett. {\bf 121}, 023401 (2018).

\bibitem{Rem13}
B. S. Rem,{\it et al}, Phys. Rev. Lett. {\bf110}, 163202 (2013).

\bibitem{Fletcher13}
R. J. Fletcher, A. L. Gaunt, N. Navon, R. P. Smith, and Z. Hadzibabic, Phys. Rev. Lett. {\bf111},125303 (2013).

\bibitem{Jin}
P. Makotyn, C. E. Klauss, D. L. Goldberger, E. A. Cornell, and D. S. Jin, Nat. Phys. {\bf 10}, 116 (2014).

\bibitem{Eigen17}
C. Eigen, {\it et al}, Phys. Rev. Lett. {\bf 119}, 250404 (2017).

\bibitem{Taylor}
J. R. Taylor, {\it Scattering Theory} (Wiley, New York, 1972), Chapter 2, 8 and 10.

\bibitem{supple}
See the supplementary material for 
I. the derivation of the identity Eq.~(\ref{theorem});  
II. the derivation of Eq.~(\ref{phik}); 
III. the simplified equations for the Bogoliubov ansatz; 
IV. the discussion of the oscillatory solution of the Bogoliubov equations. 

\bibitem{note}
We can always change $\psi(p)$ in an infinitesimal small neighborhood of $p=0$ to make $\psi(p=0)=0$, in order to satisfy the requirement of being an odd function.  


\bibitem{Cheng}
J. Hu, L. Feng, Z. Zhang, C. Chin, Nature Physics {\bf15}, 785 (2019).

\bibitem{Meera}
A. Mu\~noz de las Heras, M. M. Parish, F. M. Marchetti, Phys. Rev. A {\bf99}, 023623 (2019).


\end{thebibliography}

\end{document}


\title{Supplementary material: \\
Universal Dynamics of a Degenerate Bose Gas Quenched to Unitarity}

\author{Chao Gao}
\email{gaochao@zjnu.edu.cn}
\affiliation{Department of Physics, Zhejiang Normal University, Jinhua, 321004, China}
\author{Mingyuan Sun}
\affiliation{Institute for Advanced Study, Tsinghua University, Beijing, 100084, China}
\author{Peng Zhang}
\affiliation{Department of Physics, Renmin University of China, Beijing, 100872,
China}
\affiliation{Beijing Key Laboratory of Opto-electronic Functional Materials \&
Micro-nano Devices, 100872 (Renmin Univeristy of China)}
\author{Hui Zhai}
\email{hzhai@tsinghua.edu.cn}
\affiliation{Institute for Advanced Study, Tsinghua University, Beijing, 100084, China}

\date{\today}
\maketitle

\section{I. Proof of Eq. (4)}

In this section we prove the relation Eq.~(4) in the main text, i.e., the relation 
\begin{eqnarray}
n_{{\bf k}}(t\to +\infty )=|\langle {\bf k}|\hat{\Omega }^{(-)\dagger }|\phi ^i\rangle |^2=|\langle{\bf k}^{(-)}|\phi ^i\rangle |^2.\label{s1}
\end{eqnarray}
As shown in our main text, here $|\phi^i\rangle$ is the initial state, which is a normalized wave packet, $|{\bf k}^{(-)}\rangle$ is the inward scattering wave function defined in Eq. (5) of our main text, 
\begin{eqnarray}
n_{{\bf k}}(t)=|\langle {\bf k}|e^{-\frac{i}{\hbar}\hat{H}t}|\phi^{i}\rangle|^2=|\langle {\bf k}|e^{\frac{i}{\hbar}\hat{H}_0 t}e^{-\frac{i}{\hbar}\hat{H}t}|\phi^{i}\rangle|^2\label{s2}
\end{eqnarray}
is the momentum distribution defined in Eq.~(2) of our main text, and
\begin{eqnarray}
\hat{\Omega }^{(-)}=\lim_{t\to +\infty } e^{\frac{i}{\hbar } \hat{H} t}e^{-\frac{i}{\hbar } \hat{H}_0t}
\end{eqnarray}   
is the so-called M{\o}ller operator.

First,
we emphasis that here in the unitary case, the Hamiltonian $H$ does not support any bound state. As shown in the end of Sec. 2-d of Ref.~\cite{Taylor}, in this case the M{\o}ller operator $\hat{\Omega }^{(-)}$ is an unitary operator. Namely, $\hat{\Omega }^{(-)}$ is an invertible map from the Hilbert space to itself, and satisfies $\hat{\Omega }^{(-)-1}=\hat{\Omega }^{(-)\dagger}$. Thus, for any normalizable state $|\phi^i\rangle$, we can define  another normalizable state $|\tilde{\phi}^i\rangle$ as
\begin{eqnarray}
|\tilde{\phi}^i\rangle\equiv\hat{\Omega }^{(-)-1}|\phi^i\rangle=\hat{\Omega }^{(-)\dagger}|\phi^i\rangle.\label{psitilde}
\end{eqnarray}
It is clear that $|\tilde{\phi}^i\rangle$ satisfies
\begin{eqnarray}
\hat{\Omega }^{(-)}|\tilde{\phi}^i\rangle=|\phi^i\rangle,
\end{eqnarray}
or
\begin{eqnarray}
\lim_{t\rightarrow+\infty}\left\Vert e^{\frac{i}{\hbar } \hat{H} t}e^{-\frac{i}{\hbar } \hat{H}_0t}
|\tilde{\phi}^i\rangle-|\phi^i\rangle \right\Vert=0,\label{ss5bb}
\end{eqnarray}
where $\left\Vert|\phi\rangle\right\Vert=\sqrt{\langle\phi|\phi\rangle}$. Notice that Eq. (\ref{ss5bb}) yields
\begin{eqnarray}
\lim_{t\rightarrow+\infty}\left\Vert |\tilde{\phi}^i\rangle-e^{\frac{i}{\hbar } \hat{H}_0 t}e^{-\frac{i}{\hbar } \hat{H}t}|\phi^i\rangle \right\Vert=0,\label{ss5bb}
\end{eqnarray}
or
\begin{eqnarray}
\lim_{t\to +\infty } e^{\frac{i}{\hbar } \hat{H}_0 t}e^{-\frac{i}{\hbar } \hat{H}t}|\phi ^i\rangle=|\tilde{\phi}^i\rangle.\label{ss5b}
\end{eqnarray}
Using Eq. (\ref{ss5b}) and Eq. (\ref{psitilde}), we know that for  any normalizable state $|\phi ^i\rangle$,
the limit
$\lim_{t\to +\infty } e^{\frac{i}{\hbar } \hat{H}_0 t}e^{-\frac{i}{\hbar } \hat{H}t}|\phi ^i\rangle$
does exist, and can be expressed as
\begin{eqnarray}
\lim_{t\to +\infty } e^{\frac{i}{\hbar } \hat{H}_0 t}e^{-\frac{i}{\hbar } \hat{H}t}|\phi^i\rangle=\hat{\Omega }^{(-)-1}|\phi^i\rangle=\hat{\Omega }^{(-)\dagger}|\phi^i\rangle.\label{s5}
\end{eqnarray}
Substituting Eq.~(\ref{s5}) into Eq.~(\ref{s2}), we immediately have
\begin{eqnarray}
n_{{\bf k}}(t\to +\infty )=|\langle {\bf k}|\hat{\Omega }^{(-)\dagger }|\phi ^i\rangle |^2.
\end{eqnarray}
Thus, the first equality of Eq.~(4) in the main text (i.e., Eq.~(\ref{s1})) is proved. 

Now we prove the second equality of Eq.~(4) in the main text, i.e., the result $|\langle {\bf k}|\hat{\Omega }^{(-)\dagger }|\phi ^i\rangle |^2=|\langle{\bf k}^{(-)}|\phi ^i\rangle |^2$. To this end, we first re-express $\langle {\bf k}|\hat{\Omega }^{(-)\dagger }|\phi ^i\rangle$ as
\begin{eqnarray}
\langle {\bf k}|\hat{\Omega }^{(-)\dagger }|\phi ^i\rangle=\lim_{d\rightarrow 0^+}\langle\eta_d({\bf k})|\hat{\Omega }^{(-)\dagger }|\phi ^i\rangle=\lim_{d\rightarrow 0^+}\langle\phi^i|\hat{\Omega }^{(-)}|\eta_d({\bf k})\rangle^\ast
,\label{ss11}
\end{eqnarray}
where the normalizable state $|\eta_d({\bf k})\rangle$ is defined as
\begin{eqnarray}
|\eta_d({\bf k})\rangle=\int d{\bf k}'f_d({\bf k}'-{\bf k})|{\bf k}'\rangle, 
\end{eqnarray}
with
\begin{eqnarray}
f_d({\bf k}'-{\bf k})=\frac{1}{d^3\pi^{3/2}}e^{-\frac{|{\bf k}'-{\bf k}|^2}{d^2}}.
\end{eqnarray}
Here we have used the result
\begin{eqnarray}
\delta({\bf k}'-{\bf k})=\lim_{d\rightarrow 0^+}f_d({\bf k}'-{\bf k}).
\end{eqnarray}
Furthermore, as shown in Sec. 8-c, 10-a and 10-b of Ref.~\cite{Taylor}, 
for the normalizable state $|\eta_d\rangle$
we also have
\begin{eqnarray}
\hat{\Omega }^{(-) }|\eta_d({\bf k})\rangle=\int d{\bf k}'f_d({\bf k}'-{\bf k})|{\bf k}'^{(-)}\rangle,\label{s9}
\end{eqnarray}
with the scattering state $|{\bf k}'^{(-)}\rangle$ being defined in Eq. (5) of our main text. Thus, substituting Eq. (\ref{s9}) into Eq. (\ref{ss11}), we obtain
\begin{eqnarray}
\langle {\bf k}|\hat{\Omega }^{(-)\dagger }|\phi ^i\rangle =\langle\phi ^i|{\bf k}^{(-)}\rangle^\ast,
\end{eqnarray}
  which gives
\begin{eqnarray}
|\langle {\bf k}|\hat{\Omega }^{(-)\dagger }|\phi ^i\rangle |^2=|\langle{\bf k}^{(-)}|\phi ^i\rangle |^2.
\end{eqnarray}
Thus, the second equality of Eq.~(4) in the main text (i.e., Eq.~(\ref{s1})) is proved.

\section{II. Proof of Eq. (10)}

Now we prove Eq. (10) of our main text. First, substituting Eq. (6) of our main text into Eq. (9) of our main text, and using the fact that $\psi({\bf p})$ is independent of the direction of ${\bf p}$ (i.e., $\psi({\bf p})=\psi(p)$), we obtain
\begin{eqnarray}
\phi^i(k)&=&\frac{1}{(2\pi)^3}\int d^3{\bf r}\left[\int d^3{\bf p}e^{-i{\bf k}\cdot{\bf r}}\psi(p)\left(e^{i{\bf p}\cdot{\bf r}}+\frac{1}{ip}\frac{e^{-ipr}}{r}\right)\right]\nonumber\\
&=&\frac{2}{i\pi}\int_0^{+\infty}dr\left[\int_0^{+\infty}dpF(p,r)\right],\label{ss}
\end{eqnarray}
with the function $F(p,r)$ being defined as
\begin{eqnarray}
F(p,r)&=&\left(\frac pk\right)\psi(p)\sin(kr)\cos(pr).\label{ss2}
\end{eqnarray}
Furthermore, since the integration $\int_0^{+\infty}dr\left[\int_0^{+\infty}dpF(p,r)\right]$ converges, we have
\begin{eqnarray}
\phi^i(k)
&=&\lim_{\epsilon\rightarrow 0^+}\frac{2}{i\pi}\int_0^{+\infty}dr e^{-\epsilon r}\left[\int_0^{+\infty}dpF(p,r)\right].\label{ss3}
\end{eqnarray}
Using this result we can exchange the integrations $\int dr$ and $\int dp$, and have
\begin{eqnarray}
\phi^i(k)
&=&\lim_{\epsilon\rightarrow 0^+}\frac{2}{i\pi}\int_0^{+\infty}dp \left[\int_0^{+\infty}dr F(p,r)e^{-\epsilon r}\right]\nonumber\\
&=&\frac{1}{2 \pi }\left(-\frac{i}{k}\right)\lim_{\epsilon \to 0^+} \int _0^{+\infty }dpp \psi(p)
\left[\sum _{\sigma ,\sigma '=\pm } \frac{\sigma }{p+\sigma  (k+i \epsilon  \sigma ')}\right].\nonumber\\
\label{ss3}
\end{eqnarray}
This is Eq. (10) of our main text.
 
\section{III. simplified equations for the Bogoliubov ansatz}
In following we present a simplified form of the coupled equations of motion, Eq.(14), in the main text:

\begin{equation}\ba
i\dot{\tilde{g}}_0&=A_1\tilde{g}_0+\int _0^{\infty }dk A_2(k)\frac{\tilde{g}_0^*g_k+\tilde{g}_0| g_k| {}^2}{1-| g_k|{}^2},\\
i\dot{g}_p &=B_1(p)g_p+B_2(p) \left[\tilde{g}_0^2+\tilde{g}_0^{*2}g_p^2+2  | \tilde{g}_0  | {}^2g_p \right]\\
&\;\;-\frac{\pi }{2}r_0\int _0^{\infty }dk k^2 F (k r_0,p r_0 )\frac{2  | g_k  | {}^2g_p+g_k+g_k^*g_p^2}{1-  | g_k  | {}^2}.
\ea\end{equation}
Here \(\tilde{g}_0=g_0/\sqrt{N}\); \(g_k=g_{{\bf k}}\) only depends on the the module \(k= | {\bf k} |\), preserving spherical symmetry,
since we only consider \(s\)-wave interaction. Each quantity is dimensionless, with time (momentum) unit \(t_n\) (\(k_n\)), which has been set to
be unity for simplicity. The coefficient functions \(A_i\), \(B_i\), \(F\) are listed below for different interaction potentials.

For the square well potential: 
\begin{equation}\ba
&A_1=-\frac{4\gamma_\text{s}r_0}{9\pi }, \; A_2=- \frac{4\gamma_\text{s} k}{\pi } j_1 (k r_0 ),\\
&B_1=2p^2-\frac{8\gamma_\text{s}r_0}{9\pi },\; B_2=-\frac{4\gamma_\text{s}  j_1 (pr_0 )}{3\pi p},\\
&F(x,y)=\frac{8\gamma_\text{s}[y \sin (x) \cos (y)-x \cos (x) \sin (y)]}{\pi ^2 (x^3 y-x y^3 )},
\ea\end{equation}
where \(j_1(x)=[\sin (x)-x \cos (x)] /x^2 \) is the first order spherical Bessel function.

For the Gaussian potential:
\begin{equation}\ba
&A_1=-\frac{2\gamma_\text{g}r_0}{3\sqrt{\pi }},\; A_2=-\frac{2\gamma_\text{g}k^2 r_0}{\sqrt{\pi }} e^{-k^2r_0^2},\\
&B_1=2p^2-\frac{4\gamma_\text{g}r_0}{3\sqrt{\pi }},\; B_2=-\frac{2\gamma_\text{g}r_0}{3\sqrt{\pi }}e^{-p^2r_0^2}, \\
&F(x,y)=\frac{\gamma_\text{g} [e^{ -(x-y)^2 /4}-e^{ -(x+y)^2 /4} ]}{\pi ^{3/2} x y}.
\ea\end{equation}

For the Yukawa potential:
\begin{equation}\ba
&A_1=-\frac{\gamma_\text{y}r_0}{3\pi ^2},\; A_2=-\frac{\gamma_\text{y}k^2r_0}{\pi ^2 (k^2r_0^2+1 )},\\
&B_1=2p^2-\frac{2\gamma_\text{y}r_0}{3\pi ^2},\; B_2=-\frac{\gamma_\text{y} r_0}{3\pi ^2 (p^2r_0^2+1 )},\\
&F(x,y)=\frac{\gamma_\text{y}}{2\pi ^3 x y}\ln \left[\frac{(x+y)^2+1}{(x-y)^2+1} \right].
\ea\end{equation}

\section{IV. oscillatory solution of the Bogoliubov equations}

In following we discuss the oscillatory behavior of the solution of the equation-of-motion (14).
First of all, we rewrite the Bogoliubov equation-of-motion for $g_{\bf p\neq 0}$ as follows:
\be \label{EoM}
i\hbar\dot{g}_{{\bf p}}=\mc{A}_{{\bf p}}+\mc{B}_{{\bf p}} g_{\bf p}+\mc{C}_{{\bf p}} g_{\bf p}^2,
\end{equation}
where the coefficients are
\begin{equation}\ba 
\mc A_{{\bf p}} &=\frac{V({\bf p})}{L^3}g_0^2+\frac{1}{L^3}\sum_{\bf k\neq\bf0}V({\bf p-k})\frac{g_{\bf k}}{1-|g_{\bf k}|^2},\\\
\mc B_{{\bf p}} &=2\epsilon_{\bf p}+2nV(0)+
               \frac{2V(\bf p)}{L^3}|g_0|^2+\frac{1}{L^3}\sum_{\bf k\neq0}V({\bf p-k})\frac{2|g_{\bf k}|^2}{1-|g_{\bf k}|^2},\\
 \mc C_{{\bf p}}&=\mc A^*_{{\bf p}}.
\ea\ee
Let us now consider these coefficients. i) It is easy to see that $\mathcal{B}_{{\bf p}}$ is real. For short-range potentials, $V({\bf p})$ is nearly a constant for the momentum range of interests. Hence, by taking $V({\bf p-k})\approx V({\bf p})$, we have 
\begin{equation}
\frac{2V(\bf p)}{L^3}|g_0|^2+\frac{1}{L^3}\sum_{\bf k\neq0}V({\bf p-k})\frac{2|g_{\bf k}|^2}{1-|g_{\bf k}|^2}\approx 
\frac{2V(\bf p)}{L^3} \left(|g_0|^2+\sum_{\bf k\neq0}\frac{|g_{\bf k}|^2}{1-|g_{\bf k}|^2}\right)=2V({\bf p})n.
\end{equation}
Hence, we assume $\mathcal{B}({\bf p})$ is a constant independent of time. ii) Considering local two-point correlation
\begin{eqnarray}
\langle \hat{\Psi}({\bf r})\hat{\Psi}({\bf r})\rangle
=\langle \hat{a}_0\hat{a}_0+\sum_{\bf k\neq0}\hat{a}_{{\bf k}}\hat{a}_{{\bf -k}}\rangle
=\frac{1}{L^3}\left(g_0^2+\frac{g_{{\bf k}}}{1-|g_{{\bf k}}|^2}\right).
\end{eqnarray}
Note that, because of the translational symmetry, this correlation does not depend on ${\bf r}$. It is quite a common situation that, after a quench, the amplitude of local two-point correlation does not change with time after a \textit{local thermalization time}, which is of the order of a few $t_n$ in this case. Hence, again by taking $V({\bf p-k})\approx V({\bf p})$, we can write
\begin{equation}
\mc A_{{\bf p}}\approx V({\bf p})\langle \hat{\Psi}({\bf r})\hat{\Psi}({\bf r})\rangle,
\end{equation} 
which can be assumed as a constant aside from a running phase, i.e. 
\be
\mc A_{{\bf p}}\simeq {\mc A}^0_{\bf p} e^{i\omega t}, \label{ansatz}
\ee
where $ {\mc A}^0_{\bf p} $ is a constant amplitude independent of time,
and $\omega$ is a constant frequency independent both of time and momentum. 

With these two assumptions, we find that Eq.~(\ref{EoM}) can be exactly solved as
\be\label{sol}
g_{\bf p}\simeq-\frac{e^{i \omega t}} {2 \mathcal{A}^0_{{\bf p}}}
    \left(\frac{2}{\Omega_{\bf p}^{-1}+\alpha_{\bf p} e^{i  \Omega_{\bf p} t}}+\mathcal{B}_{{\bf p}}+\omega -\Omega_{\bf p} \right),
\ee
where $\Omega_{\bf p} =\sqrt{(\mathcal{B}_{{\bf p}}+\omega)^2-4 |\mathcal{A}^{0}_{\bf p}|^2}$ is the oscillation frequency of $n_{\bf k}$, and $\alpha_{\bf p}$ is a momentum-dependent parameter determined by initial condition. 

We also perform the self-consistent checks of  the assumptions and the analytical solutions. 
First, in Fig.~\ref{AB} we show $\mathcal{A}_{{\bf p}}$ from the numerical solution, and one can see that the amplitude of $\mathcal{A}_{{\bf p}}$ quickly converges to a constant, and the phase of different $\mathcal{A}_{{\bf p}}$ collapse into a single curve which linearly increases with the increasing of time, and $\mathcal{B}_{{\bf p}}$ is a constant independent of time. All these features are very much consistent with our assumptions discussed above. Secondly, we compare the analytical solution with the numerical solutions. In Fig.~\ref{nkt}, we use $g_{{\bf p}}$ from Eq.~(\ref{sol}) to obtain a momentum distribution $n_{{\bf p}}$, and compare this momentum distribution with the numerical solution of the Bogoliubov equation-of-motion. We obtain good agreement for long time dynamics, as shown in Fig.~\ref{nkt}. In this way, we have demonstrated that although the Bogoliubov equation-of-motion is highly non-linear, it does host an oscillating solution. 

\begin{figure}[t]
\begin{overpic}[width=0.4\linewidth]{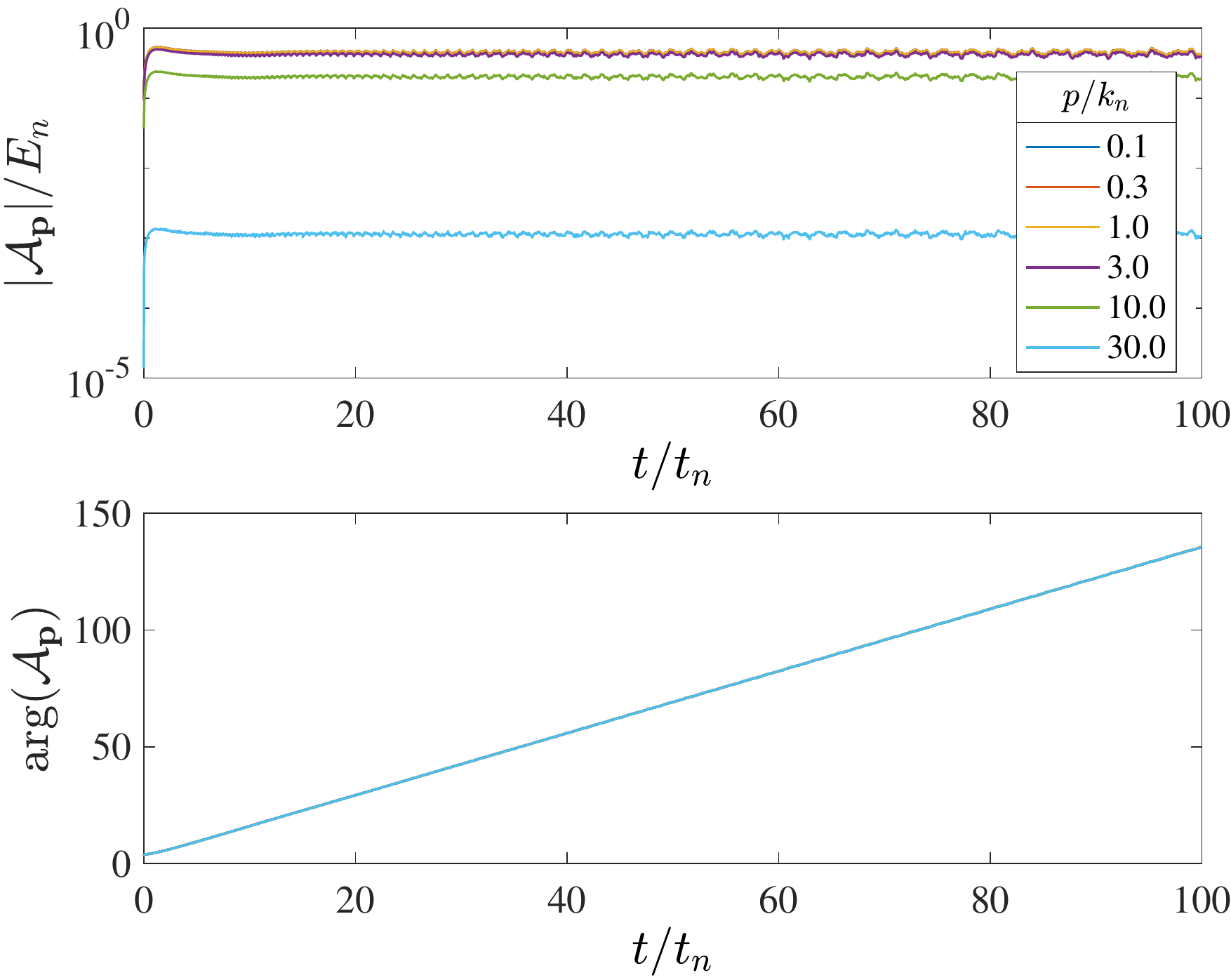}
\put(0,78){(a)}
\end{overpic}
\hspace{3mm}   \begin{overpic}[width=0.4\linewidth]{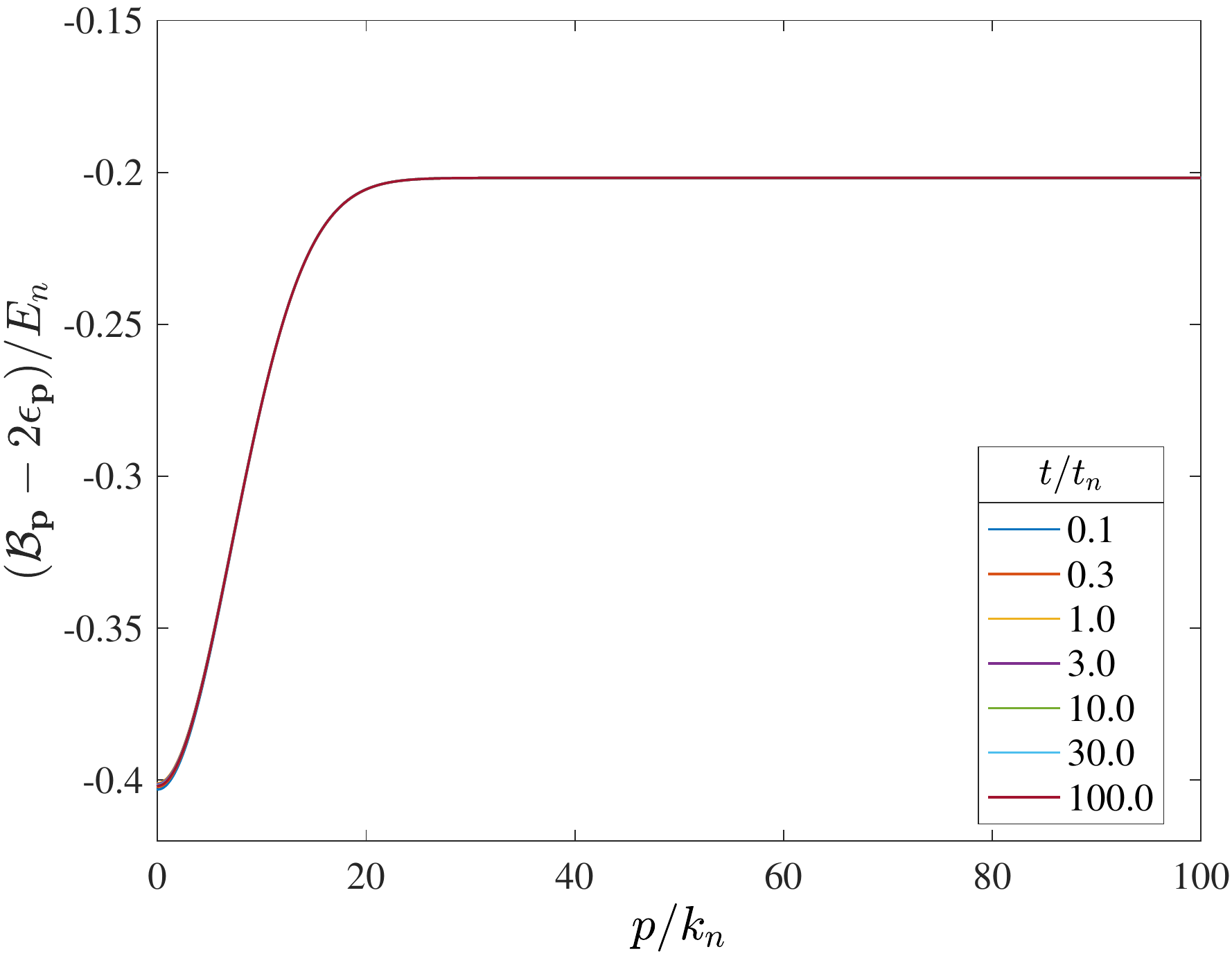}
\put(0,78){(b)}
\end{overpic}
\caption{\label{AB}
The coefficients in Eq.~(\ref{EoM}) for the case with Gaussian interaction potential.  
(a) The amplitude (top panel) and phase (bottom panel) of coefficient $\mc A_{\bf p}$ as a function of time (different colors label different momenta).
(b) The coefficient $\mc B_{\bf p}$, after extracting $2\epsilon_{\bf p}$, as a function of momentum (different colors label different times).
 }
\end{figure}

\begin{figure}[t]
\includegraphics[width=0.45\linewidth]{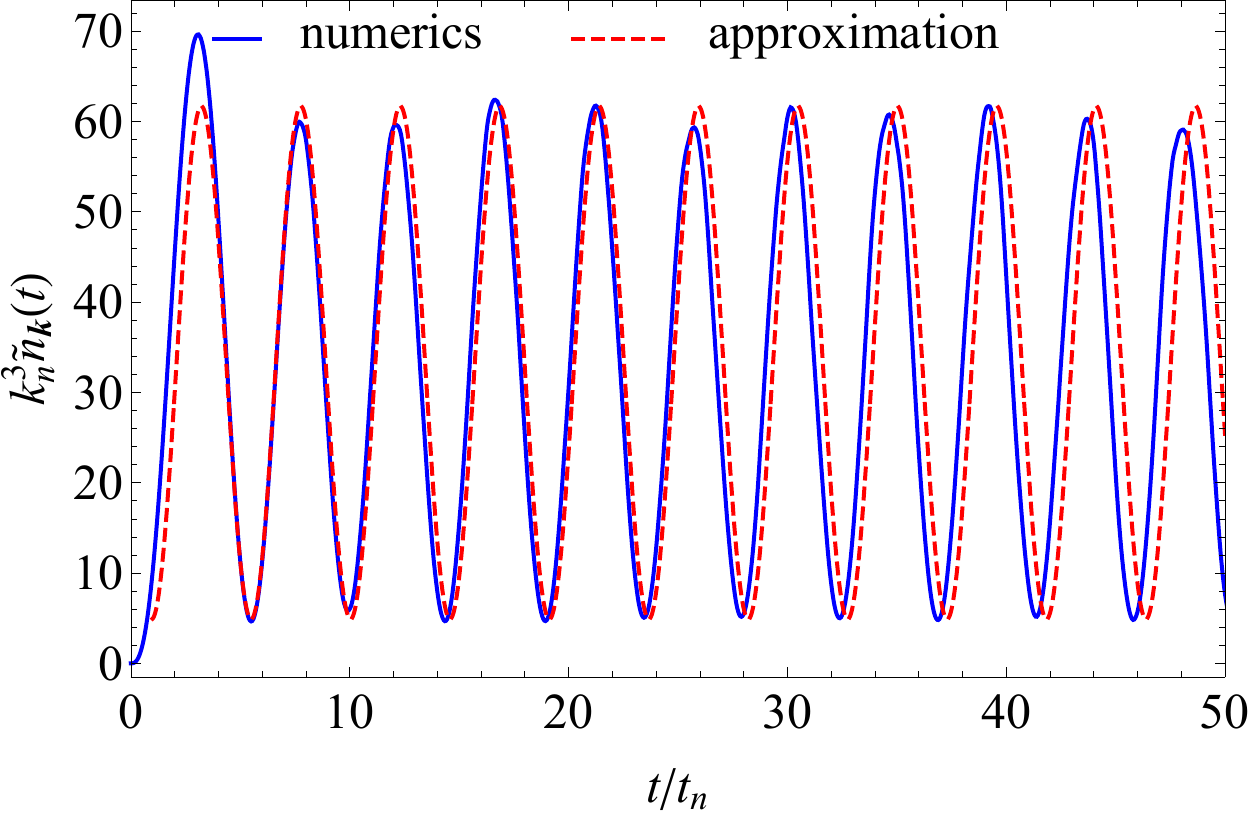}
\caption{\label{nkt}
Comparison of the numeric solution (blue solid line) and the approximated solution (red dashed line) with Eq.~(\ref{sol}) for the normalized momentum distribution as a function of time. 
Here we take a typical value of momentum, $k = 0.6 k_n$, and the interaction potential is the Gaussian potential.
The parameter $\alpha_{\bf p}$ is chosen to be $4.60e^{1.82i} t_n$.
} 
\end{figure}